\documentclass[amsmath]{elbrus6}
\usepackage{graphicx} 

\begin{document}

\newcommand{\mycmd}{My new command}

\elbrustitle{TWO-TEMPERATURE EQUATIONS OF STATE FOR D-BAND METALS IRRADIATED BY FEMOTOSECOND LASER PULSES}
\elbrusauthor{Migdal~K.\,P.$^{1}$*, Inogamov~N.\,A.$^{1,2}$, Petrov~Yu.\,V.$^{2,3}$, Zhakhovsky~V.\,V.$^1$}
\elbrusaffil{
 $^{1}$Dukhov Research Institute of Automatics, Rosatom, Moscow \\
 $^{2}$L. D. Landau Institute for Theoretical Physics, Russian Academy of Sciences, Chernogolovka \\
 $^{3}$ Moscow Institute of Physics and Technologie, Dolgoprugny}
\elbrusemail{*migdal@vniia.ru}


\begin{abstract}
The cold curves for energy and pressure of Copper, Iron, and Tantalum were obtained using methods of the density functional theory \cite{1,2}. We consider hydrostatic and uniaxial deformations in the range from double compression of the initial
volume per atom to double stretching. The presence of allotropic transformation from $\alpha$ - phase of Iron to the hexaferrum
with the growth of pressure is observed. In the case of hydrostatic deformations we also have obtained analogous cold curves,
but with non-zero electronic temperatures in the range up to 5 eV. The similar volume and electronic temperature ranges have been considered in recent works \cite{3,4}. The behaviour of electronic internal energy, pressure, and density of states was investigated in the volume and temperature ranges called above. The maximum hydrostatic strains and the types of lattice instabilities were theoretically predicted for the considered metals.
The influence of high electronic temperature on the electronic heat conductivity and electric resistivity has been provided for d-band metals by the approach based on the solution of Boltzmann kinetic equation in $\tau$-approximation \cite{5,5a,5b}. This data is compared with the results of quantum molecular dynamics for Gold \cite{6}.
\end{abstract}


\begin{figure*}
   \includegraphics[width=17cm,height=6.5cm]{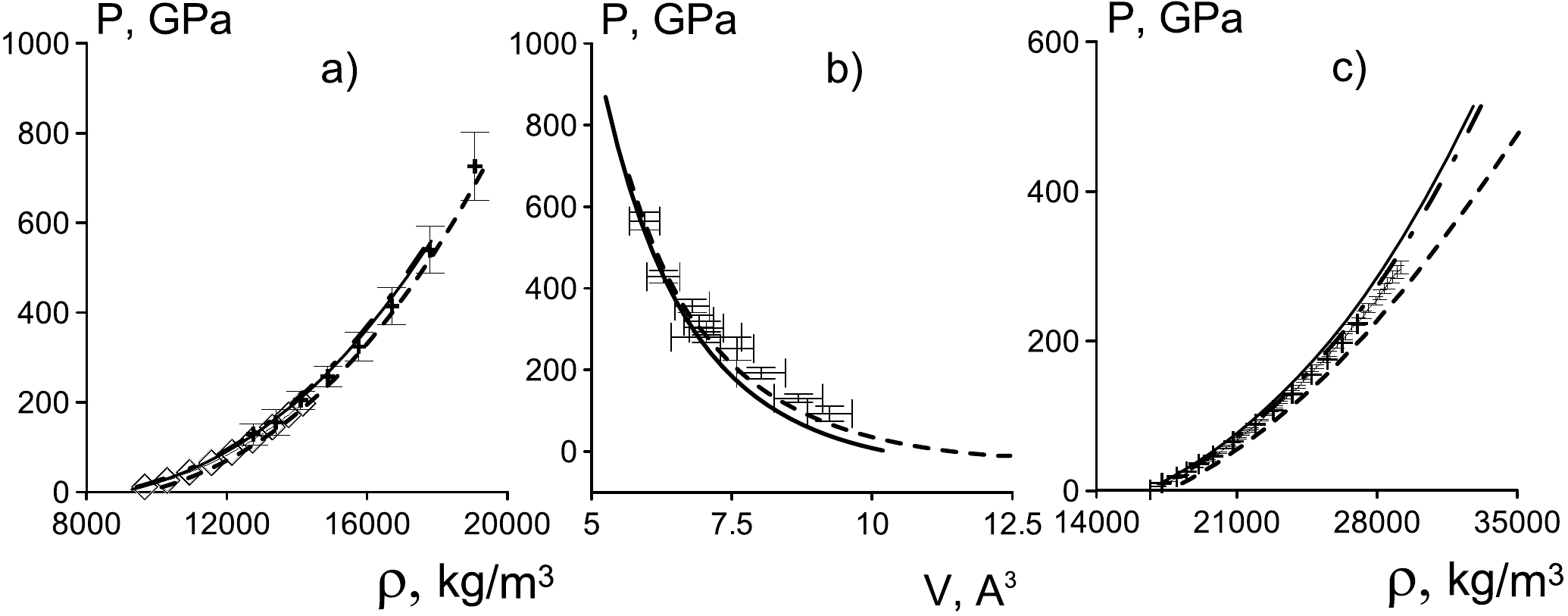}
   \caption[abs-2T-sound-nucl2]
   { \label{fig:abs-2T-sound-nucl2}
 {\it Left panel.} The calculated pressure from density for compressed Cu using LDA (dash), PBE (dash-dot), and PBEpv (solid) pseudopotentials compared with the  results of experiments \cite{7} (open diamonds) \cite{8} (thick crosses)
 {\it Middle panel.} The pressure from atomic volume for hcp (solid) and bcc (dash) compressed Fe calculated with the help of PBEpv pseudopotential and the result of recent ramp wave compression experiment \cite{13} where anharmonic approximation for Debye-Waller factor calculation have been used.
 {\it Right panel.} The calculated pressure from density for compressed Ta using LDA (dash), PBE (dash-dot), and PBEpv (solid) pseudopotentials compared with the  results of experiment \cite{7} and FP-LMTO calculation \cite{14} (thick crosses).
 }
   \end{figure*}

\ The femtosecond laser irradiation with optic wavelengths and moderate intensities acting on a metal surface is considered. With regard to metal the laser pulse energy is absorbed by free electrons on a depth of a skin layer. In this layer non-equilibrium energy distribution exists between electrons and ion lattice. In the present experimental data and theoretical calculations one have been demonstrated that a few picoseconds are required for electron-ion equilibration. The duration of electron-electron relaxation is order of magnitude 100 fs. We can consider surface layer as electron and ion subsystems, where the former obeys Fermi-Dirac distributions and has a temperature significantly greater than a temperature of the latter which is close to the metal initial temperature(~300K). This approach is valid in the case of the relation between two relaxation times mentioned above. During the two-temperature state (2T) a spatial distribution of atoms in lattice is constant because the electron-ion relaxation time is less than the typical time of acoustic unloading. Taking into account this circumstance we carried out the DFT calculation of electron thermodynamics properties for some metals in the wide range of electron temperatures and considered deformations.

\ The influence of strong hydrostatic and uniaxial strains and compressions was investigated in the cold curve calculations for metals using DFT and taking into account probabilities of allotropic transformations.

\textbf{The approach for calculations. }
The cold curves of internal energies $E$ and pressures $P$ for metals with zero lattice temperature are obtained for Cu (fcc lattice), Fe (bcc and hcp), and Ta (bcc) in the range from double compression to double expansion using the DFT code VASP \cite{1,2}. We used PAW pseudopotentials with LDA and PBE forms of exchange-correlation functional. In the latter case the closest to the conduction bands for each metal p-electron band is additionally considered. Other used parameters: plane waves cutoff is 500 eV, $21^{3}$ Monkhorst-Pack grid, and the number of free electron states is equal or greater than 20. These parameters were chosen as the result of convergence check and guarantee that convergence error is not greater than 1 \%.

\begin{figure*}
      \includegraphics[width=17cm,height=6.5cm]{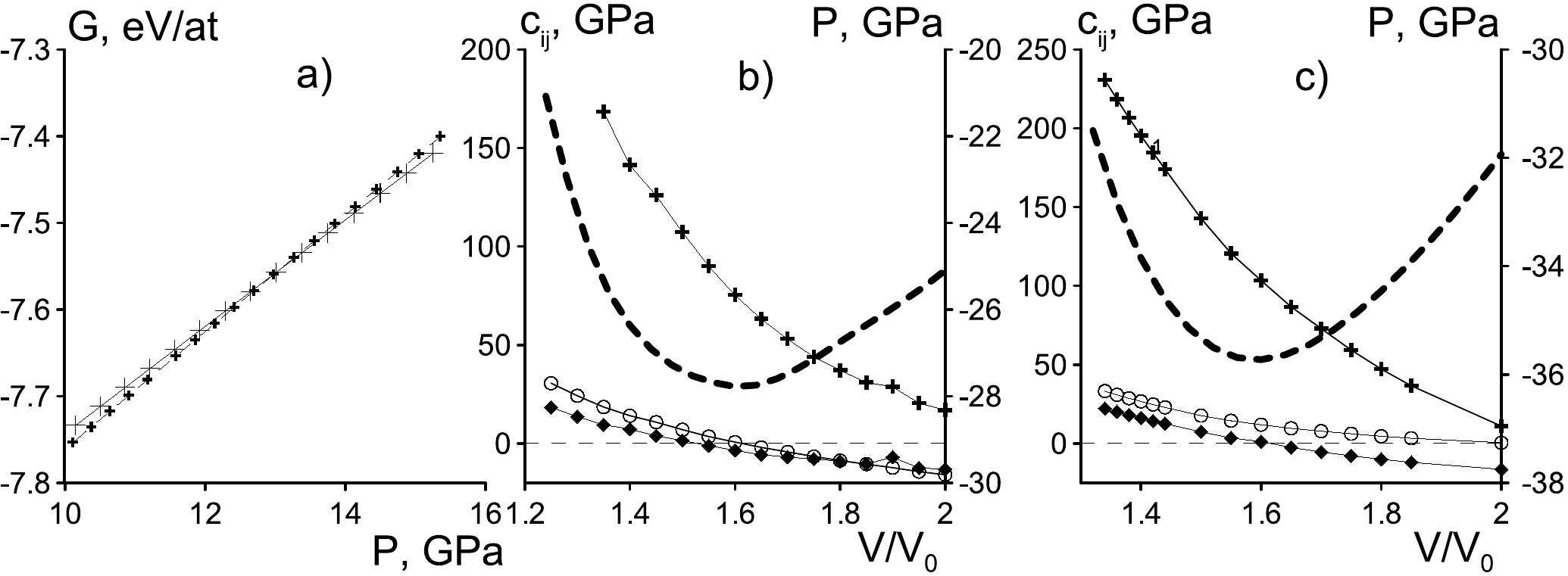}
      \caption[allotrop-latt-stability]
      { \label{fig:allotrop-latt-stability}
    {\it Left panel.} The dependence of Gibbs energy from pressure for Fe in hcp (solid line, thin crosses) and bcc (dashed line, thick crosses) phases.
    {\it Middle panel.} Left Y scale: The dependence of elastic moduli combinations - criteria of lattice stability from relative specific volume in the case of Fe. The line with thick crosses corresponds to $c_{11}+2c_{12}$ combination, open circles $\--$ $c_{11}-c_{12}$, diamonds $\--$ $c_{44}$. If the first combination is negative it means bulk instability, the second - shear instability, the third - tetragonal shear instability. Right Y scale: The dependence of pressure (dash-dot line) in bcc Fe from relative specific volume in the range corresponding to expansion.
    {\it Right panel.} The same values as in the middle panel for Ta.
           }
      \end{figure*}

\ The validity of the value of expanded volume where metal lattice becomes unstable was approved by the DFT calculation of elastic moduli. The cutoffs were enlarged up to 600 eV (Cu), 800 eV (Fe), 650 eV (Ta). The obtained values for elastic moduli for some metal specific volumes at hydrostatic expansion allow to determine the type of lattice instability which is the reason of metal breakdown. The presence of hot electrons have been took into account with the use of smearing value and the Fermi-Dirac smearing was used. We carried out the calculation of 2T equations of state in the range from double compression to double expansion for electron temperatures up to 55 000K.

\textbf{The results and comparison with experimental data.}
On the Fig. 1a the obtained cold curve for Cu is compared with the data of experiments \cite{7,8} in the case of compression. As we can see all curves are close to each other and the PBEpv curve is in better agreement with experimental data than other curves. The PBEpv curve for hcp (solid line) and bcc Fe (dashed line) is compared with the experimental data \cite{13} obtained by ramp wave compression (RWC) method. This method allow to obtain the values of pressure for solid matter which cannot be achieved using previously used technique. In \cite{13} the new record for solid Fe was received - 560 GPa. It is necessary to notice that the dependence of pressure from compression obtained in this work not pure experimental curve but the result of experimental data treatment with the help of QMD simulation and Debye model. The analogous comparison for Ta is shown in Fig 1a where the experimental and FP-LMTO calculation data are in good agreement with our results (mainly PBEpv curve).

\begin{figure*}
      \includegraphics[width=17cm,height=6.5cm]{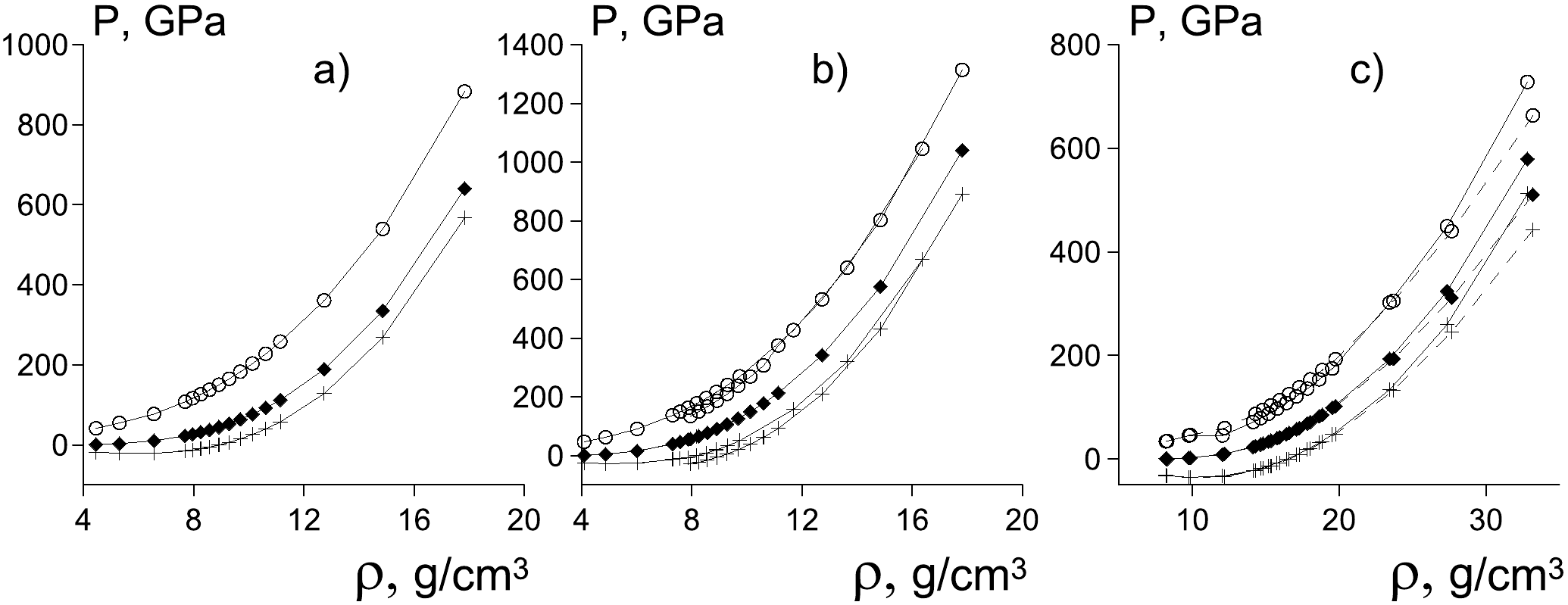}
      \caption[allotrop-latt-stability4]
      { \label{fig:allotrop-latt-stability4}
      The dependence of pressure from density at hydrostatic deformations obtained with the use of PBEpv pseudopotential for fixed values of electron temperature $T_e$=1000 K (thin crosses), 25000 K(diamonds), 55000 R (open circles).
    {\it Left panel.} The case of Cu.
    {\it Middle panel.} The case of bcc and hcp Fe. For the values of $T_e$ 1000 K and 55 000 K the hcp curves is always low than bcc. For the value of $T_e$ the difference between two curves are negligibly small.
    {\it Right panel.} The case of Ta. The dashed lines corresponds to dependences obtained with the use of PBE pseudopotential.
           }
      \end{figure*}

\ The dependences of Gibbs energy from pressure for bcc and hcp lattices of Fe are shown on the Fig 2a in the range of $\alpha-\varepsilon$ allotropic transformation search. As the result of this search the pressure of bcc-hcp transformation is found with taking into account of uncertainties of $G$ due to convergence error and finite step of the grid of pressure values. The found values is $14 \frac{+5}{-3}$ GPa (PBE) and $13 \frac{+5}{-2}$ (PBEpv) and in good agreement with the results of known experimental data \cite{9}. This received values are obtained only after addition calculation on more denser gred (100 points of $V/V_0$ in the compression range from 0.9 to 1 $V/V_0$).

\begin{figure*}
      \includegraphics[width=12cm,height=6.5cm]{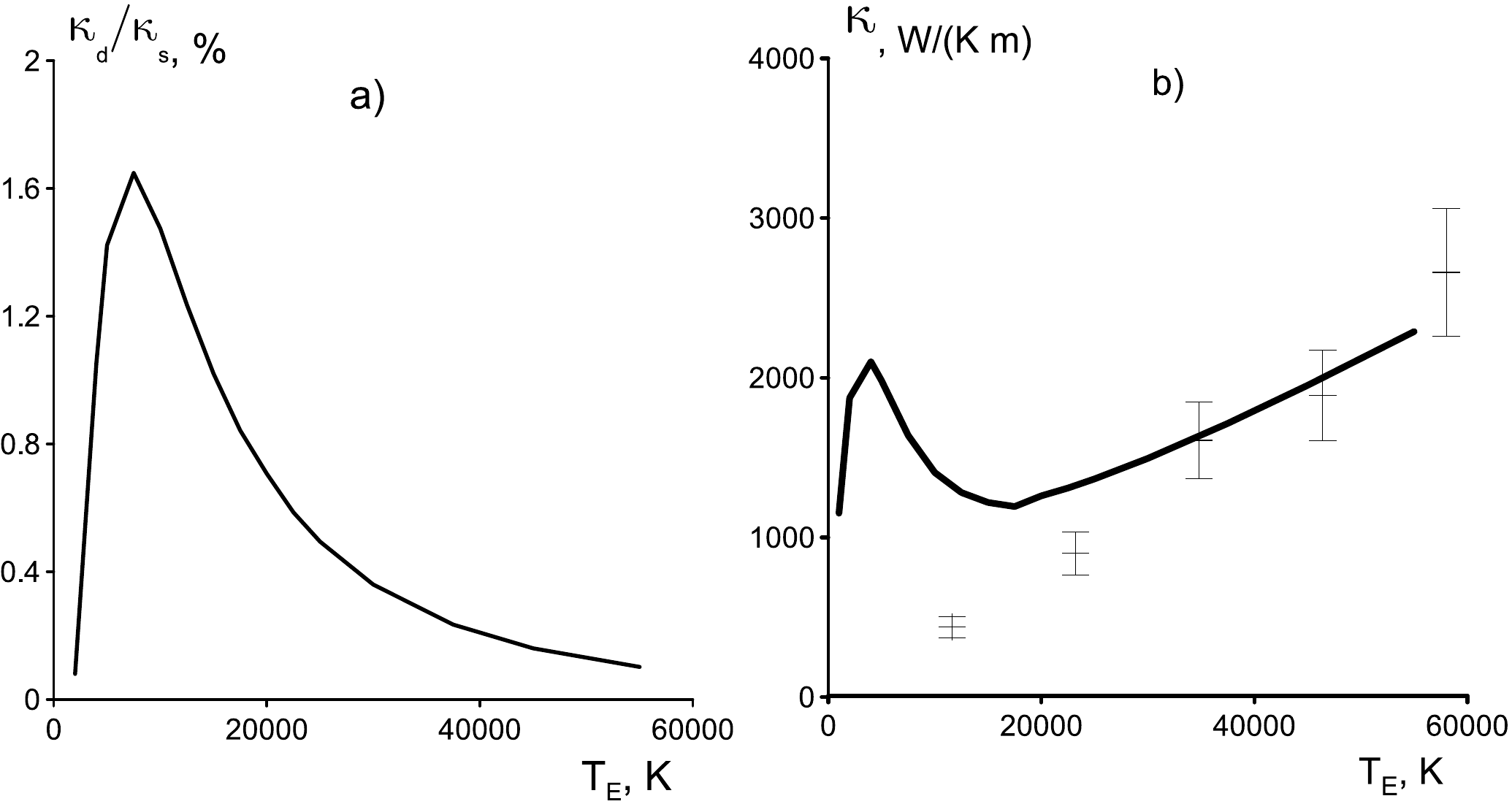}
      \caption[abs-2T-sound-nucl]
      { \label{fig:abs-2T-sound-nucl1}
    {\it Left panel.}  The dependence from electron temperature of the overestimated d-electron thermal conductivity divided by s-electron thermal conductivity for Cu.
    {\it Right panel.} The dependence from electron temperature of the s-electron thermal conductivity for Au and quantum molecular dynamics simulation performed in \cite{6}.

       }
      \end{figure*}

\ On the Fig 2b and 2c the values for Fe and Ta of elastic moduli combinations which negative values (even if one of them) determine the area of lattice instability in the case of hydrostatic expansion. The minimum value of pressure is achieved at the specific volume where the lattice becomes unstable. As we can see for both metals the results for cold curves and  more accurate elastic moduli calculations are in good agreement because the growth of pressure with volume is obvious case of unphysical behaviour at expansion.

\ The results for 2T EoS for Cu, Fe, and Ta are shown on Fig 3 as the dependencies of pressure from density at some values of electron temperature. The influence of additionally considered p-electron band in PBE pseudopotential is negligible for most cases (cannot be shown on this figure) and only for strong compression of Ta (Fig 3c) the difference caused by this band is sufficient to take it into account. The positive shift of pressure due to the growth of electron temperature is more noticeable for Cu than for transition metals. In the case of Fe the described above bcc-hcp transition cannot be reliably found at the demonstrated values for electron temperature 25000K and 55000K. We can explain such behavior of Fe as the consequence of decrease of lattice influence on metal state when the electron subsystem energy becomes sufficiently high.

       \begin{figure*}
            \includegraphics[width=12cm,height=6.5cm]{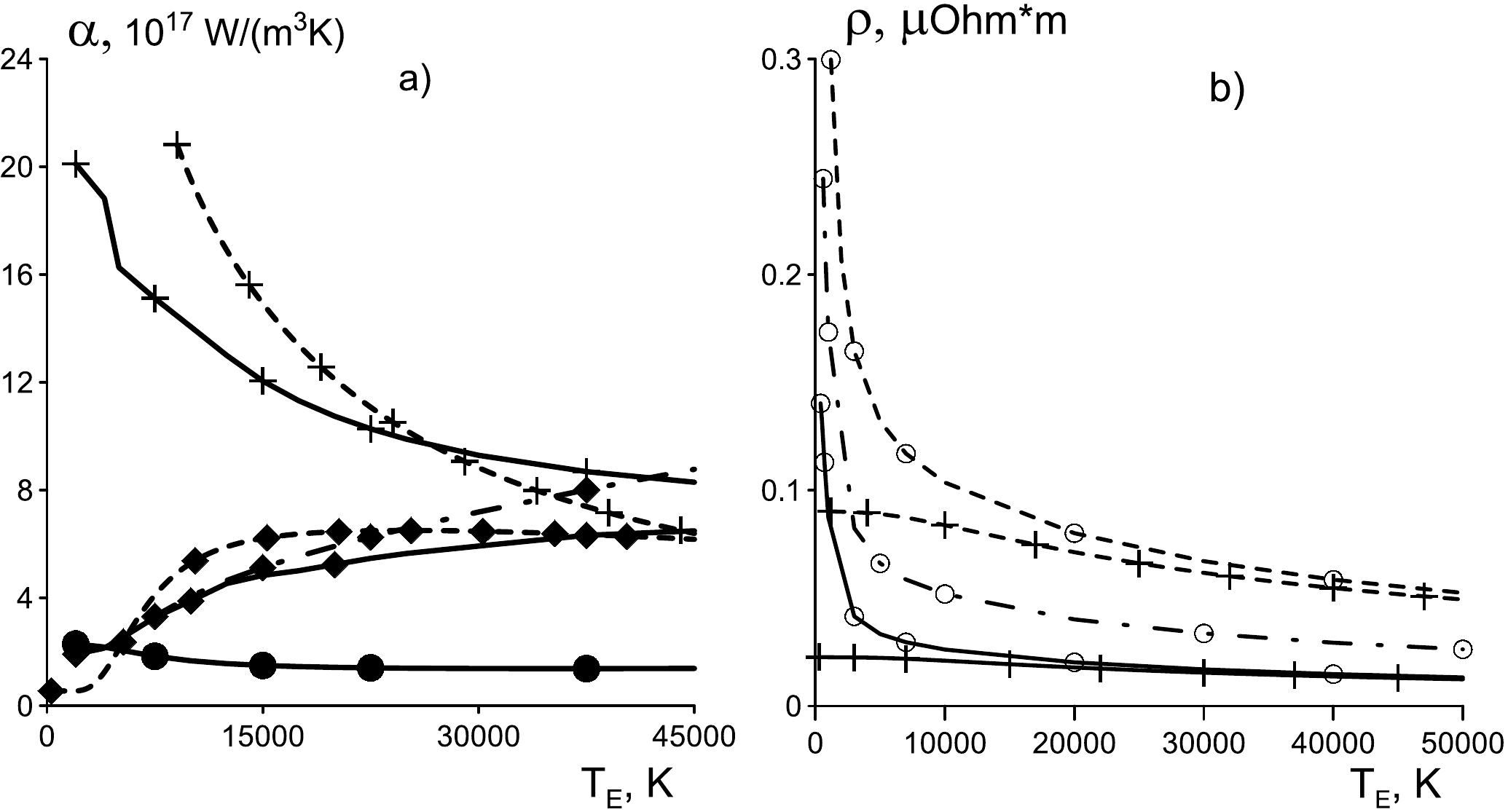}
            \caption[abs-2T-sound-nucl3]
            { \label{fig:abs-2T-sound-nucl3}
         {\it Left panel.} The dependence from electron temperature of electron-ion coupling for Cu (diamonds), Fe (thick crosses) and Ta (open circles). The obtained with taking into account of $g(T_e)$ dependence data for Cu and Fe are shown by solid lines. The previous result for Cu without the consideration of electron temperature influence on DOS is shown by dash-dotted line and the results from web resource \cite{15} are shown by dashed line.
          {\it Right panel.}  Resistivities $\sigma$ of Au (thick crosses) and Ni (open circles) as functions of electron temperature $T_e.$
            The functions $\sigma_{Au}(T_e)$ are given at two fixed values of lattice temperature (300 K - solid line, 1200 - dashed line) and $\sigma_{Fe}(T_e)$ at three fixed ion temperatures $T_i$=300 K(solid), 600 K(dash-dot) and 1200 K (dash).
                     }
            \end{figure*}

\textbf{Electronic transport coefficients calculation.}
The calculation of the electron conductivity coefficient was carried out with the use of approach \cite{5,10} which is based on the result for Boltzmann kinetic equation for free electrons in the relaxation time approximation.
On Fig 4a the data for $\kappa_{2T}$ obtained with taking into account of the electronic spectra dependence from electron temperature and found previously cite{10} are represented. Even if we neglect the frequency of  d-electron - ion collisions  so the result for d-electron conductivity $\kappa_{2T,d}$ is overestimated, the value of
$\frac{\kappa_{2T,d}}{\kappa_{2T,s}}$ is less than 0.02 on the full electron temperature range. The calculated by DFT partial electron densities of states allow us to check the statement that the s-electron impact in electronic conductivity is dominated. On Fig 4b the comparison between the results for $\kappa_{2T}$ obtained in our two parabolic approximation \cite{10} and QMD calculations based on Kubo-Greenwood approach \cite{6} is provided for the case of Au. We can see on Fig 3a that two areas of electron temperatures exist. On the first area where $T_e < 30 000K$ our result has a peak at 5000 K and slightly decrease while the QMD values increase monotonously with $T_e$. The second area starts from $T_e>30000K$ and there the considered results are in good agreement. Authors supposed that the peak of $\kappa_{2T}$ is the consequence of weak d-band influence on collision frequencies at temperatures which less than the absolute value of d-band edge $\varepsilon_{2,Au}=-1.7 eV$.

\ Taking into consideration the electron temperature influence on electron density of states causes the noticeable decrease of electron-ion coupling in the case of Cu, as we can see on Fig 5a. The represented on Fig 5a values of coupling $\alpha$ was obtained using Kaganov-Lifshitz-Tanatarov approach \cite{10,11}.

\ Also the contribution of electron-phonon interaction in energy and charge transfer was carried out at high electron temperatures. The determination of this contribution makes possible to obtain the accurate analytical expression for electron resistivity of metal with hot electrons \cite{12}.

\ The phonon spectra was approximated by modified Debye approach where the deviation from linear acoustic behavior at large values of phonon momentum is taking into account. For two-band metals all possible processes for s- and d-band electrons are considered and the conditions on the values of electron and phonon momenta have been taken into account during Boltzmann kinetic equation solutions integration. These solutions were obtained for electrons interacting with outer electric field in relaxation time approximation. The electric conductivity is determined with the use of Ohm law.

\ As a result the curves of electric conductivity $\sigma$ dependence from $T_e$ was calculated fro Au and Ni. On the Fig 5b one demonstrated that the value $\sigma$ is almost not depend from $T_e$ for noble metals(Au). But the same value for Ni demonstrates steep decreasing at temperatures is order of magnitude of melting temperature for Ni. It is worth to notice that the demonstrated dependence was not taking into account in the previous known to the authors works.

This work was supported by Russian Science Foundation grant 14-19-01599.


\end{document}